\documentclass[12pt]{article}

\title{
\hfill{\normalsize ULB/229/CQ/00/6}\\
\vspace{1cm}
Para, pseudo, and orthosupersymmetric quantum mechanics and their bosonization}

\author{C. Quesne\thanks{E-mail: cquesne@ulb.ac.be}\\
{\small \sl PNTPM, Universit\'e Libre de Bruxelles, Campus de la Plaine
CP229,}\\
{\small \sl Boulevard du Triomphe, B-1050 Brussels, Belgium}}

\date{ }
\begin{document}
\baselineskip=20pt plus 1pt minus 1pt
\maketitle

\begin{abstract}
We consider the problem of bosonizing supersymmetric quantum mechanics
(SSQM) and some of its variants, i.e., of realizing them in terms of only
boson-like operators without fermion-like ones. In the SSQM case, this is
realized
in terms of the generators of the Calogero-Vasiliev algebra (also termed
deformed
Heisenberg algebra with reflection). In that of the SSQM variants, this is
done by
considering generalizations of the latter algebra, namely the
$C_{\lambda}$-extended oscillator algebras, where $C_{\lambda}$ is the cyclic
group of order~$\lambda$.
\end{abstract}

\section{Introduction}

Supersymmetry has established an elegant symmetry between bosons and fermions
and is one of the cornerstones of modern theoretical physics. Its application to
quantum mechanics has provided a powerful method of generating solvable quantum
mechanical models. On the other hand, exotic quantum statistics have received
considerable attention due to their possible relevance to the fractional quantum
Hall effect and anyon superconductivity.\par
%
%
By combining both concepts within the framework of quantum mechanics, one gets
variants of SSQM: paraSSQM~\cite{cq:rubakov,cq:khare93a,cq:beckers90},
pseudoSSQM~\cite{cq:beckers95a,cq:beckers95b}, and
orthoSSQM~\cite{cq:khare93b}. They can be realized in terms of bosons and
parafermions~\cite{cq:ohnuki},
pseudofermions~\cite{cq:beckers95a,cq:beckers95b},
or orthofermions~\cite{cq:mishra}, respectively.\par
%
%
By using the Calogero-Vasiliev algebra~\cite{cq:vasiliev}, Plyushchay
showed~\cite{cq:plyu} that SSQM can be described in terms of only boson-like
operators without fermion-like ones (see also \cite{cq:beckers97}).\par
%
%
In the present communication, we shall consider generalizations of the
Calogero-Vasiliev algebra, namely the $C_{\lambda}$-extended oscillator algebras
(where $C_{\lambda} = \mbox{Z}_{\lambda}$ is the cyclic group of order
$\lambda$)~\cite{cq:cq98,cq:cq99,cq:cq00}. We shall show that they have some
interesting applications to variants of SSQM~\cite{cq:cq98,cq:cq00}, as
they provide
a bosonization of the latter analogous to that obtained by Plyushchay for
SSQM.\par
%
%
\section{Generalized deformed and $\cal G$-extended oscillator algebras}

The generalized deformed oscillator algebras (GDOAs) (see e.g.\ Refs.\
\cite{cq:cq95,cq:cq96} and references quoted therein) arose from successive
generalizations of the Arik-Coon~\cite{cq:arik} and
Biedenharn-Macfarlane~\cite{cq:biedenharn,cq:macfarlane}
$q$-oscillators. Such algebras, denoted by ${\cal A}_q(G(N))$, are
generated by the
unit, creation, annihilation, and number operators $I$, $a^{\dagger}$, $a$, $N$,
satisfying the Hermiticity conditions $\left(a^{\dagger}\right)^{\dagger} = a$,
$N^{\dagger} = N$, and the commutation relations
\begin{equation}
  \left[N, a^{\dagger}\right] = a^{\dagger}, \qquad [N, a] = - a, \qquad
  \left[a, a^{\dagger}\right]_q \equiv a a^{\dagger} - q a^{\dagger} a = G(N),
\end{equation}
where $q$ is some real number and $G(N)$ is some Hermitian, analytic
function.\par
%
%
On the other hand, $\cal G$-extended oscillator algebras, where $\cal G$ is some
finite group, appeared in connection  with $n$-particle integrable models.
For the
Calogero model~\cite{cq:calogero71}, for instance, $\cal G$ is the
symmetric group $S_n$~\cite{cq:poly,cq:brink}.\par
%
%
{}For two particles, the $S_2$-extended oscillator algebra ${\cal
A}^{(2)}_{\kappa}$, where $S_2 = \{\, I, K \mid K^2 = I \,\}$, is generated
by the
operators $I$, $a^{\dagger}$, $a$, $N$, $K$, subject to the Hermiticity
conditions
$\left(a^{\dagger}\right)^{\dagger} = a$, $N^{\dagger} = N$, $K^{\dagger} =
K^{-1}$,
and the relations
\begin{eqnarray}
  \left[N, a^{\dagger}\right] & = & a^{\dagger}, \qquad [N, K] = 0, \qquad
K^2 = I,
          \nonumber \\
  \left[a, a^{\dagger}\right] & = & I + \kappa K \qquad (\kappa \in {\mbox R}),
  \qquad a^{\dagger} K = - K a^{\dagger},
\end{eqnarray}
together with their Hermitian conjugates.\par
%
%
When the $S_2$ generator $K$ is realized in terms of the Klein operator
$(-1)^N$,
${\cal A}^{(2)}_{\kappa}$ becomes a GDOA characterized by $q=1$ and $G(N) = I +
\kappa (-1)^N$, and known as the Calogero-Vasiliev oscillator
algebra~\cite{cq:vasiliev}.\par
%
%
The operator $K$ may be alternatively considered as the generator of the cyclic
group $C_2$ of order two, since the latter is isomorphic to $S_2$. By replacing
$C_2$ by the cyclic group of order $\lambda$, $C_{\lambda} = \{\, I, T,
T^2, \ldots,
T^{\lambda-1} \mid T^{\lambda} = I \,\}$, one then gets a new class of $\cal
G$-extended oscillator algebras~\cite{cq:cq98,cq:cq99,cq:cq00},
generalizing that
describing the two-particle Calogero model.\par
%
%
\section{\boldmath $C_{\lambda}$-extended oscillator algebras}

Let us consider the algebras generated by the operators $I$, $a^{\dagger}$,
$a$, $N$,
$T$, satisfying the Hermiticity conditions
$\left(a^{\dagger}\right)^{\dagger} = a$,
$N^{\dagger} = N$, $T^{\dagger} = T^{-1}$, and the relations
\begin{eqnarray}
  \left[N, a^{\dagger}\right] & = & a^{\dagger}, \qquad [N, T] = 0, \qquad
T^{\lambda} =
            I, \nonumber \\
  \left[a, a^{\dagger}\right] & = & I + \sum_{\mu=1}^{\lambda-1} \kappa_{\mu}
            T^{\mu}, \qquad a^{\dagger} T = e^{-{\rm i}2\pi/\lambda}\, T
a^{\dagger},
\label{eq:alg-def1}
\end{eqnarray}
together with their Hermitian conjugates~\cite{cq:cq98}. Here $T$ is the
generator
of (a unitary representation of) the cyclic group $C_{\lambda}$ (where
$\lambda \in
\{2, 3, 4, \ldots\}$), and $\kappa_{\mu}$, $\mu = 1$, 2,
$\ldots$,~$\lambda-1$, are
some complex parameters restricted by the conditions $\kappa_{\mu}^* =
\kappa_{\lambda - \mu}$ (so that there remain altogether $\lambda-1$
independent real parameters).\par
%
%
$C_{\lambda}$ has $\lambda$ inequivalent, one-dimensional matrix unitary
irreducible representations (unirreps) $\Gamma^{\mu}$, $\mu = 0$, 1,
$\ldots$,~$\lambda-1$, which are such that $\Gamma^{\mu}\left(T^{\nu}\right) =
\exp({\rm i}2\pi \mu \nu/\lambda)$ for any $\nu = 0$, 1, $\ldots$,~$\lambda-1$.
The projection operator on the carrier space of~$\Gamma^{\mu}$ may be written as
\begin{equation}
  P_{\mu} = \frac{1}{\lambda} \sum_{\nu=0}^{\lambda-1}
  e^{-{\rm i}2\pi \mu\nu/\lambda}\, T^{\nu},
\end{equation}
and conversely $T^{\nu}$, $\nu=0$, 1, $\ldots$,~$\lambda-1$, may be expressed in
terms of the $P_{\mu}$'s as
\begin{equation}
  T^{\nu} = \sum_{\mu=0}^{\lambda-1}  e^{{\rm i}2\pi \mu\nu/\lambda} P_{\mu}.
\end{equation}
\par
%
%
The algebra defining relations~(\ref{eq:alg-def1}) may therefore be rewritten in
terms of $I$, $a^{\dagger}$, $a$, $N$, and~$P_{\mu}^{\vphantom{\dagger}} =
P_{\mu}^{\dagger}$, $\mu=0$, 1, $\ldots$,~$\lambda-1$, as
\begin{eqnarray}
  \left[N, a^{\dagger}\right] & = & a^{\dagger}, \qquad \left[N,
P_{\mu}\right] = 0,
            \qquad \sum_{\mu=0}^{\lambda-1} P_{\mu} = I, \nonumber \\
  \left[a, a^{\dagger}\right] & = & I + \sum_{\mu=0}^{\lambda-1} \alpha_{\mu}
            P_{\mu}, \qquad a^{\dagger} P_{\mu} = P_{\mu+1}\, a^{\dagger},
\qquad
            P_{\mu} P_{\nu} = \delta_{\mu,\nu} P_{\mu},  \label{eq:alg-def2}
\end{eqnarray}
where we use the convention $P_{\mu'} = P_{\mu}$ if $\mu' - \mu = 0\, {\rm
mod}\,
\lambda$ (and similarly for other operators or parameters indexed by $\mu$,
$\mu'$). Equation~(\ref{eq:alg-def2}) depends upon $\lambda$ real parameters
$\alpha_{\mu} = \sum_{\nu=1}^{\lambda-1} \exp({\rm i}2\pi \mu\nu/\lambda)
\kappa_{\nu}$, $\mu=0$, 1, $\ldots$,~$\lambda-1$, restricted by the condition
$\sum_{\mu=0}^{\lambda-1} \alpha_{\mu} = 0$. Hence, we may eliminate one of
them, for instance $\alpha_{\lambda-1}$, and denote $C_{\lambda}$-extended
oscillator algebras by ${\cal A}^{(\lambda)}_{\alpha_0 \alpha_1 \ldots
\alpha_{\lambda-2}}$.\par
%
%
The cyclic group generator $T$ and the projection operators $P_{\mu}$ can be
realized in terms of $N$ as
\begin{equation}
  T = e^{{\rm i}2\pi N/\lambda}, \qquad P_{\mu} = \frac{1}{\lambda}
  \sum_{\nu=0}^{\lambda-1} e^{{\rm i}2\pi \nu (N-\mu)/\lambda}, \qquad \mu
= 0, 1,
  \ldots, \lambda-1,    \label{eq:N-realize}
\end{equation}
respectively. With such a choice, ${\cal A}^{(\lambda)}_{\alpha_0 \alpha_1
\ldots
\alpha_{\lambda-2}}$\ becomes a GDOA, ${\cal A}^{(\lambda)}(G(N))$,
characterized
by
$q=1$ and $G(N) = I + \sum_{\mu=0}^{\lambda-1} \alpha_{\mu} P_{\mu}$, where
$P_{\mu}$ is given in Eq.~(\ref{eq:N-realize}).\par
%
%
{}For any GDOA ${\cal A}_q(G(N))$, one may define a so-called structure
function~$F(N)$, which is the solution of the difference equation $F(N+1) -
q F(N) =
G(N)$, such that $F(0) = 0$~\cite{cq:cq95}. For ${\cal
A}^{(\lambda)}(G(N))$, we find
\begin{equation}
  F(N) = N + \sum_{\mu=0}^{\lambda-1} \beta_{\mu} P_{\mu}, \quad \beta_0
  \equiv 0, \quad \beta_{\mu} \equiv \sum_{\nu=0}^{\mu-1} \alpha_{\nu} \quad
  (\mu =1, 2, \ldots, \lambda-1).
\end{equation}
\par
%
%
At this point, it is worth noting that for $\lambda=2$, we obtain $T=K$,
$P_0 = (I +
K)/2$, $P_1 = (I - K)/2$, and $\kappa_1 = \kappa_1^* = \alpha_0 = - \alpha_1 =
\kappa$, so that ${\cal A}^{(2)}_{\alpha_0}$ coincides with the $S_2$-extended
oscillator algebra ${\cal A}^{(2)}_{\kappa}$ and ${\cal A}^{(2)}(G(N))$ with the
Calogero-Vasiliev algebra.\par
%
%
In Ref.~\cite{cq:cq00}, it was shown that ${\cal A}^{(\lambda)}(G(N))$ (and more
generally ${\cal A}^{(\lambda)}_{\alpha_0 \alpha_1 \ldots \alpha_{\lambda-2}}$)
has only two different types of unirreps: infinite-dimensional bounded from below
unirreps and finite-dimensional ones. Among the former, there is the so-called
bosonic Fock space representation, wherein $a^{\dagger} a = F(N)$ and $a
a^{\dagger}
= F(N+1)$. Its carrier space $\cal F$ is spanned by the
eigenvectors~$|n\rangle$ of
the number operator~$N$, corresponding to the eigenvalues $n=0$, 1, 2,~$\ldots$,
where $|0\rangle$ is a vacuum state, i.e., $a |0\rangle = N|0\rangle = 0$ and
$P_{\mu} |0\rangle = \delta_{\mu,0} |0\rangle$. The eigenvectors can be
written as
\begin{equation}
  |n\rangle = {\cal N}_n^{-1/2} \left(a^{\dagger}\right)^n |0\rangle,
\qquad n = 0, 1, 2,
  \ldots,  \label{eq:vectors}
\end{equation}
where ${\cal N}_n = \prod_{i=1}^n F(i)$. The creation and annihilation
operators act
upon~$|n\rangle$ in the usual way, i.e.,
\begin{equation}
  a^{\dagger} |n\rangle = \sqrt{F(n+1)}\, |n+1\rangle, \qquad a |n\rangle =
\sqrt{F(n)}
  \, |n-1\rangle,
\end{equation}
while $P_{\mu}$ projects on the $\mu$th component ${\cal F}_{\mu} \equiv
\{\, |k\lambda + \mu\rangle \mid k = 0, 1, 2, \ldots\,\}$ of the
${\rm Z}_{\lambda}$-graded Fock space ${\cal F} = \sum_{\mu=0}^{\lambda-1}
\oplus {\cal F}_{\mu}$. It is obvious that such a bosonic Fock space
representation
exists if and only if $F(\mu) > 0$ for $\mu=1$, 2, $\ldots$,~$\lambda-1$. This
gives the following restrictions on the algebra parameters~$\alpha_{\mu}$,
\begin{equation}
  \sum_{\nu=0}^{\mu-1} \alpha_{\nu} > - \mu, \qquad \mu = 1, 2, \ldots,
\lambda-1.
  \label{eq:cond-Fock}
\end{equation}
\par
%
%
In the bosonic Fock space representation, one may consider the bosonic
oscillator
Hamiltonian, defined as usual by
\begin{equation}
  H_0 \equiv {\textstyle{1\over 2}} \left\{a, a^{\dagger}\right\}.
\label{eq:H_0}
\end{equation}
It can be rewritten as
\begin{equation}
  H_0 = a^{\dagger} a + \frac{1}{2} \left(I + \sum_{\mu=0}^{\lambda-1}
\alpha_{\mu}
  P_{\mu}\right) = N + \frac{1}{2} I + \sum_{\mu=0}^{\lambda-1} \gamma_{\mu}
  P_{\mu},
\end{equation}
where $\gamma_0 \equiv \frac{1}{2} \alpha_0$ and $\gamma_{\mu} \equiv
\sum_{\nu=0}^{\mu-1} \alpha_{\nu} + \frac{1}{2} \alpha_{\mu}$ for $\mu = 1$, 2,
\ldots,~$\lambda-1$.\par
%
%
The eigenvectors of $H_0$ are the states~$|n\rangle = |k \lambda + \mu\rangle$,
defined in Eq.~(\ref{eq:vectors}), and their eigenvalues are given by
\begin{equation}
  E_{k\lambda+\mu} = k\lambda + \mu + \gamma_{\mu} + {\textstyle{1\over2}},
  \qquad k = 0, 1, 2, \ldots, \qquad \mu = 0, 1, \ldots, \lambda-1.
\end{equation}
In each ${\cal F}_{\mu}$ subspace of the ${\rm Z}_{\lambda}$-graded Fock
space~$\cal F$, the spectrum of~$H_0$ is therefore harmonic, but the $\lambda$
infinite sets of equally spaced energy levels, corresponding to $\mu=0$, 1,
$\ldots$,~$\lambda-1$, may be shifted with respect to each other by some
amounts depending upon the algebra parameters $\alpha_0$, $\alpha_1$,
$\ldots$,~$\alpha_{\lambda-2}$, through their linear combinations
$\gamma_{\mu}$, $\mu=0$, 1, $\ldots$,~$\lambda-1$.\par
%
%
{}For the Calogero-Vasiliev oscillator, i.e., for $\lambda=2$, the relation
$\gamma_0 = \gamma_1 = \kappa/2$ implies that the spectrum is very simple and
coincides with that of a shifted harmonic oscillator. For $\lambda\ge 3$,
however,
it has a much richer structure. According to the parameter values, it may be
nondegenerate, or may exhibit some ($\nu+1$)-fold degeneracies above some energy
eigenvalue, where $\nu$ may take any value in the set $\{1, 2, \ldots,
\lambda-1\}$.
In Ref.~\cite{cq:cq99}, the complete classification of nondegenerate,
twofold and
threefold degenerate spectra was obtained for $\lambda=3$ in terms of $\alpha_0$
and $\alpha_1$.\par
%
%
In the remaining part of this communication, we will show that the bosonic Fock
space representation of ${\cal A}^{(\lambda)}(G(N))$ and the corresponding bosonic
oscillator Hamiltonian $H_0$ have some useful applications to variants of
SSQM.\par
%
%
\section{\boldmath Application to parasupersymmetric quantum mechanics of order
$p$}

In SSQM with two supercharges, the supersymmetric Hamiltonian $\cal H$ and the
supercharges $Q^{\dagger}$, $Q = \left(Q^{\dagger}\right)^{\dagger}$,
satisfy the
sqm(2) superalgebra, defined by the relations
\begin{equation}
  Q^2 = 0, \qquad [{\cal H}, Q] = 0, \qquad \left\{Q, Q^{\dagger}\right\} =
{\cal H},
  \label{eq:SSQM}
\end{equation}
together with their Hermitian conjugates. Such a superalgebra is
most often realized in terms of mutually commuting boson and fermion
operators.\par
%
%
Plyushchay~\cite{cq:plyu}, however, showed that it can alternatively be
realized in
terms of only boson-like operators, namely the generators of the
Calogero-Vasiliev
algebra ${\cal A}^{(2)}(G(N))$ (see also Ref.~\cite{cq:beckers97}). The SSQM
bosonization can be performed in two different ways, by choosing either $Q =
a^{\dagger} P_1$ (so that ${\cal H} = H_0 - \frac{1}{2}(K + \kappa)$) or $Q =
a^{\dagger} P_0$ (so that ${\cal H} = H_0 + \frac{1}{2}(K + \kappa)$). The first
choice corresponds to unbroken SSQM (all the excited states are twofold
degenerate
while the ground state is nondegenerate and at vanishing energy), and the second
choice describes broken SSQM (all the states are twofold degenerate and at
positive
energy).\par
%
%
SSQM was generalized to parasupersymmetric quantum mechanics (PS\-SQM) of
order two by Rubakov and Spiridonov~\cite{cq:rubakov}, and later on to PSSQM of
arbitrary order $p$ by Khare~\cite{cq:khare93a}. In the latter case,
Eq.~(\ref{eq:SSQM}) is replaced by
\[
Q^{p+1} = 0 \qquad ({\rm with\ } Q^p \ne 0),
\]
\[
  [{\cal H}, Q] = 0,
\]
\begin{equation}
  Q^p Q^{\dagger} + Q^{p-1} Q^{\dagger} Q + \cdots + Q Q^{\dagger} Q^{p-1} +
  Q^{\dagger} Q^p = 2p Q^{p-1} {\cal H}, \label{eq:PSSQM}
\end{equation}
and is retrieved in the case where $p=1$. The parasupercharges $Q$,
$Q^{\dagger}$,
and the parasupersymmetric Hamiltonian $\cal H$ are usually realized in terms of
mutually commuting boson and parafermion operators.\par
%
%
A property of PSSQM of order $p$ is that the spectrum of $\cal H$ is
($p+1$)-fold
degenerate above the ($p-1$)th energy level. This fact and Plyushchay's
results for
$p=1$ hint at a possibility of representing $\cal H$ as a linear
combination of the
bosonic oscillator Hamiltonian $H_0$ associated with ${\cal
A}^{(p+1)}(G(N))$ and
some projection operators.\par
%
%
In Ref.~\cite{cq:cq00} (see also Ref.~\cite{cq:cq98}), it was proved that PSSQM
of order $p$ can indeed be bosonized in terms of the generators of ${\cal
A}^{(p+1)}(G(N))$ for any allowed (i.e., satisfying
Eq.~(\ref{eq:cond-Fock})) values of
the algebra parameters $\alpha_0$, $\alpha_1$, \ldots,~$\alpha_{p-1}$. For
such a
purpose, ans\" atze of the type
\begin{equation}
  Q = \sum_{\nu=0}^p \sigma_{\nu} a^{\dagger} P_{\nu}, \qquad {\cal H} = H_0 +
  {\textstyle{1\over 2}} \sum_{\nu=0}^p r_{\nu} P_{\nu},
\end{equation}
were chosen. Here $\sigma_{\nu}$ and $r_{\nu}$ are some complex and real
constants, respectively, to be determined in such a way that
Eq.~(\ref{eq:PSSQM})
is fulfilled. It was found that there are $p+1$ families of solutions,
which may be
distinguished by an index $\mu \in \{0, 1, \ldots, p\}$ and from which one may
choose the following representative solutions
\begin{eqnarray}
  Q_{\mu} & = & \sqrt{2} \sum_{\nu=1}^p a^{\dagger} P_{\mu+\nu}, \nonumber\\
  {\cal H}_{\mu} & = &  N + {\textstyle{1\over 2}} (2\gamma_{\mu+2} +
  r_{\mu+2} - 2p + 3) I + \sum_{\nu=1}^p (p + 1 - \nu) P_{\mu+\nu},
  \label{eq:PSSQM-sol}
\end{eqnarray}
where
\begin{equation}
  r_{\mu+2} = \frac{1}{p} \left[(p-2) \alpha_{\mu+2} + 2 \sum_{\nu=3}^p
(p-\nu+1)
  \alpha_{\mu+\nu} + p (p-2)\right].
\end{equation}
\par
%
%
The eigenvectors of ${\cal H}_{\mu}$ are the states~(\ref{eq:vectors}) and the
corresponding eigenvalues are easily found. All the energy levels are equally
spaced. For $\mu=0$, PSSQM is unbroken, otherwise it is broken with a
($\mu+1$)-fold degenerate ground state. All the excited states are ($p+1$)-fold
degenerate. For $\mu=0$, 1, \ldots,~$p-2$, the ground state energy may be
positive,
null, or negative depending on the parameters, whereas for $\mu = p-1$ or
$p$, it is
always positive.\par
%
%
Khare~\cite{cq:khare93a} showed that in PSSQM of order $p$, $\cal H$ has in fact
$2p$ (and not only two) conserved parasupercharges, as well as $p$ bosonic
constants. In other words, there exist $p$ independent operators $Q_r$,
$r=1$, 2,
\ldots,~$p$, satisfying with $\cal H$ the set of equations~(\ref{eq:PSSQM}), and
$p$ other independent operators $I_t$, $t=2$, 3, \ldots,~$p+1$, commuting with
$\cal H$, as well as among themselves. In Ref.~\cite{cq:cq00}, a
realization of all
such operators was obtained in terms of the ${\cal A}^{(p+1)}(G(N))$
generators.\par
%
%
As a final point, let us note that there exists an alternative approach to
PSSQM of
order $p$, which was proposed by Beckers and Debergh~\cite{cq:beckers90}, and
wherein the multilinear relation in Eq.~(\ref{eq:PSSQM}) is replaced by the
cubic
equation
\begin{equation}
  \left[Q, \left[Q^{\dagger}, Q\right] \right] = 2Q {\cal H}.  \label{eq:cubic}
\end{equation}
In Ref.~\cite{cq:cq98}, it was proved that for $p=2$, this PSSQM algebra
can only be
realized by those ${\cal A}^{(3)}(G(N))$ algebras that simultaneously bosonize
Rubakov-Spiridonov-Khare PSSQM algebra.\par
%
%
\section{Application to pseudosupersymmetric quantum mechanics}

Pseudosupersymmetric quantum mechanics (pseudoSSQM) was introduced by
Beckers, Debergh, and Nikitin~\cite{cq:beckers95a,cq:beckers95b} in a study of
relativistic vector mesons interacting with an external constant magnetic
field. In
the nonrelativistic limit, their theory leads to a pseudosupersymmetric
oscillator
Hamiltonian, which can be realized in terms of mutually commuting boson and
pseudofermion operators, where the latter are intermediate between standard
fermion and $p=2$ parafermion operators.\par
%
%
It is then possible to formulate a
pseudoSSQM~\cite{cq:beckers95a,cq:beckers95b},
characterized by a pseudosupersymmetric Hamiltonian $\cal H$ and
pseudosupercharge operators
$Q$,
$Q^{\dagger}$, satisfying the relations
\begin{equation}
  Q^2 = 0, \qquad [{\cal H}, Q] = 0, \qquad Q Q^{\dagger} Q = 4 c^2 Q {\cal H},
  \label{eq:pseudoSSQM}
\end{equation}
and their Hermitian conjugates, where $c$ is some real constant. The first two
relations in Eq.~(\ref{eq:pseudoSSQM}) are the same as those occurring in SSQM,
whereas the third one is similar to the multilinear relation valid in PSSQM
of order
two. Actually, for $c=1$ or 1/2, it is compatible with Eq.~(\ref{eq:PSSQM}) or
(\ref{eq:cubic}), respectively.\par
%
%
In Ref.~\cite{cq:cq00}, it was proved that pseudoSSQM can be bosonized in two
different ways in terms of the generators of ${\cal A}^{(3)}(G(N))$ for any
allowed
values of the parameters $\alpha_0$, $\alpha_1$. This time, the ans\" atze
\begin{equation}
  Q = \sum_{\nu=0}^2 \left(\xi_{\nu} a + \eta_{\nu} a^{\dagger}\right) P_{\nu},
  \qquad {\cal H} = H_0 + {\textstyle{1\over 2}} \sum_{\nu=0}^2 r_{\nu}
P_{\nu},
\end{equation}
were chosen, and the complex constants $\xi_{\nu}$, $\eta_{\nu}$, and
the real ones $r_{\nu}$ were determined in such a way that
Eq.~(\ref{eq:pseudoSSQM}) is fulfilled.\par
%
%
The first type of bosonization corresponds to three families of two-parameter
solutions, labelled by an index $\mu \in \{0, 1, 2\}$,
\begin{eqnarray}
  Q_{\mu}(\eta_{\mu+2}, \varphi) & = & \left(\eta_{\mu+2} a^{\dagger} +
e^{{\rm i}
          \varphi}\sqrt{4 c^2 - \eta_{\mu+2}^2}\, a\right) P_{\mu+2},
\nonumber \\
  {\cal H}_{\mu}(\eta_{\mu+2}) & = & N + {\textstyle{1\over 2}} (2
          \gamma_{\mu+2} + r_{\mu+2} - 1) I + 2 P_{\mu+1} + P_{\mu+2},
          \label{eq:pseudoSSQM-sol}
\end{eqnarray}
where $0 < \eta_{\mu+2} < 2 |c|$, $0 \le \varphi < 2\pi$, and
\begin{equation}
  r_{\mu+2} = \frac{1}{2c^2} (1 + \alpha_{\mu+2}) \left(|\eta_{\mu+2}|^2 - 2
  c^2\right).
\end{equation}
Choosing for instance $\eta_{\mu+2} = \sqrt{2} |c|$, and $\varphi = 0$, hence
$r_{\mu+2} = 0$ (producing an overall shift of the spectrum), leads to
\begin{eqnarray}
  Q_{\mu} & = & c \sqrt{2} \left(a^{\dagger} + a\right) P_{\mu+2}, \nonumber \\
  {\cal H}_{\mu} & = & N + {\textstyle{1\over 2}} (2 \gamma_{\mu+2} - 1) I
  + 2 P_{\mu+1} + P_{\mu+2}.  \label{eq:pseudoSSQM-solbis}
\end{eqnarray}
A comparison between Eq.~(\ref{eq:pseudoSSQM-sol}) or
(\ref{eq:pseudoSSQM-solbis}) and Eq.~(\ref{eq:PSSQM-sol}) shows that the
pseudosupersymmetric and $p=2$ parasupersymmetric Hamiltonians coincide, but
that the corresponding charges are of course different. The conclusions
relative to
the spectrum and the ground state energy are therefore the same as in Sec.~4.
\par
%
%
The second type of bosonization corresponds to three families of one-parameter
solutions, again labelled by an index $\mu \in \{0, 1, 2\}$,
\begin{eqnarray}
  Q_{\mu} & = & 2 |c| a P_{\mu+2}, \nonumber \\
  {\cal H}_{\mu}(r_{\mu}) & = & N + {\textstyle{1\over 2}} (2 \gamma_{\mu+2}
          - \alpha_{\mu+2}) I + {\textstyle{1\over 2}} (1 - \alpha_{\mu+1} +
          \alpha_{\mu+2} + r_{\mu}) P_{\mu} \nonumber \\
  && \mbox{} + P_{\mu+1},
\end{eqnarray}
where $r_{\mu} \in {\rm R}$ changes the Hamiltonian spectrum in a significant
way. The levels are indeed equally spaced if and only if $r_{\mu} =
(\alpha_{\mu+1} - \alpha_{\mu+2} + 3)\, {\rm mod}\, 6$. If $r_{\mu}$ is small
enough, the ground state is nondegenerate, and its energy is negative for
$\mu=1$,
or may have any sign for $\mu = 0$ or~2. On the contrary, if $r_{\mu}$ is large
enough, the ground state remains nondegenerate with a vanishing energy in the
former case, while it becomes twofold degenerate with a positive energy in the
latter. For some intermediate $r_{\mu}$ value, one gets a two or threefold
degenerate ground state with a vanishing or positive energy, respectively.\par
%
%
\section{Application to orthosupersymmetric quantum mechanics of order two}

Mishra and Rajasekaran~\cite{cq:mishra} introduced order-$p$ orthofermion
operators by replacing the Pauli exclusion principle by a more stringent one: an
orbital state shall not contain more than one particle, whatever be the spin
direction. The wave function is thus antisymmetric in spatial indices alone
with the
order of the spin indices frozen.\par
%
%
Khare, Mishra, and Rajasekaran~\cite{cq:khare93b} then developed
orthosupersymmetric quantum mechanics (OSSQM) of arbitrary order $p$ by
combining boson operators with orthofermion ones, for which the spatial indices
are ignored. OSSQM is formulated in terms of an orthosupersymmetric Hamiltonian
$\cal H$, and $2p$ orthosupercharge operators $Q_r$, $Q_r^{\dagger}$, $r =
1$, 2,
\ldots,~$p$, satisfying the relations
\begin{equation}
  Q_r Q_s = 0, \qquad [{\cal H}, Q_r] = 0, \qquad Q_r Q_s^{\dagger} +
\delta_{r,s}
  \sum_{t=1}^p Q_t^{\dagger} Q_t = 2 \delta_{r,s} {\cal H},  \label{eq:OSSQM}
\end{equation}
and their Hermitian conjugates, where $r$ and $s$ run over 1, 2, \ldots,~$p$.\par
%
%
In Ref.~\cite{cq:cq00}, it was proved that OSSQM of order two can be
bosonized in
terms of the generators of some well-chosen ${\cal A}^{(3)}(G(N))$ algebras. As
ans\" atze, the expressions
\begin{eqnarray}
  Q_1 & = & \sum_{\nu=0}^2 \left(\xi_{\nu} a + \eta_{\nu} a^{\dagger}\right)
         P_{\nu}, \qquad Q_2 = \sum_{\nu=0}^2 \left(\zeta_{\nu} a + \rho_{\nu}
         a^{\dagger}\right) P_{\nu}, \nonumber \\
 {\cal H} & = & H_0 + {\textstyle{1\over 2}} \sum_{\nu=0}^2 r_{\nu} P_{\nu},
\end{eqnarray}
were used, and the complex constants $\xi_{\nu}$, $\eta_{\nu}$, $\zeta_{\nu}$,
$\rho_{\nu}$, and the real ones $r_{\nu}$ were determined in such a way that
Eq.~(\ref{eq:OSSQM}) is fulfilled. There exist two families of two-parameter
solutions, labelled by $\mu \in
\{0 ,1\}$,
\begin{eqnarray}
  Q_{1,\mu}(\xi_{\mu+2}, \varphi) & = & \xi_{\mu+2} a P_{\mu+2} + e^{{\rm i}
         \varphi} \sqrt{2 - \xi_{\mu+2}^2}\, a^{\dagger} P_{\mu}, \nonumber \\
  Q_{2,\mu}(\xi_{\mu+2}, \varphi) & = & - e^{-{\rm i} \varphi} \sqrt{2 -
         \xi_{\mu+2}^2}\, a P_{\mu+2} + \xi_{\mu+2} a^{\dagger} P_{\mu},
         \nonumber\\
  {\cal H}_{\mu} & = & N + {\textstyle{1\over 2}} (2 \gamma_{\mu+1} - 1) I
         + 2 P_{\mu} + P_{\mu+1},  \label{eq:OSSQM-sol}
\end{eqnarray}
where $0 < \xi_{\mu+2} \le \sqrt{2}$ and $0 \le \varphi <2\pi$, provided the
algebra parameter $\alpha_{\mu+1}$ is taken as $\alpha_{\mu+1} = -1$. As a
matter
of fact, the absence of a third family of solutions corresponding to
$\mu=2$ comes
from the incompatibility of this condition (i.e., $\alpha_0 = -1$) with
conditions~(\ref{eq:cond-Fock}).\par
%
%
The orthosupersymmetric Hamiltonian $\cal H$ in Eq.~(\ref{eq:OSSQM-sol}) is
independent of the parameters $\xi_{\mu+2}$, $\varphi$. All the levels of its
spectrum are equally spaced. For $\mu=0$, OSSQM is broken: the levels are
threefold
degenerate, and the ground state energy is positive. On the contrary, for
$\mu=1$,
OSSQM is unbroken: only the excited states are threefold degenerate, while the
nondegenerate ground state has a vanishing energy. Such results agree with the
general conclusions of Ref.\ \cite{cq:khare93b}.\par
%
%
{}For $p$ values greater than two, the OSSQM algebra~(\ref{eq:OSSQM}) becomes
rather complicated because the number of equations to be fulfilled increases
considerably. A glance at the 18 independent conditions for $p=3$ led to the
conclusion that the ${\cal A}^{(4)}(G(N))$ algebra is not rich enough to contain
operators satisfying Eq.~(\ref{eq:OSSQM}). Contrary to what happens for PSSQM,
for OSSQM the $p=2$ case is therefore not representative of the general one.\par
%
%
\section{Conclusion}

In this communication, we showed that the $S_2$-extended oscillator algebra,
which was introduced in connection with the two-particle Calogero model, can be
extended to the whole class of $C_{\lambda}$-extended oscillator algebras ${\cal
A}^{(\lambda)}_{\alpha_0 \alpha_1 \ldots \alpha_{\lambda-2}}$, where $\lambda
\in \{2,3, \ldots\}$, and $\alpha_0$, $\alpha_1$,
\ldots,~$\alpha_{\lambda-2}$ are
some real parameters. In the same way, the GDOA realization of the former, known
as the Calogero-Vasiliev algebra, is generalized to a class of GDOAs ${\cal
A}^{(\lambda)}(G(N))$, where $\lambda \in \{2,3, \ldots\}$, for which one
can define
a bosonic oscillator Hamiltonian $H_0$, acting in the bosonic Fock space
representation.\par
%
%
{}For $\lambda \ge 3$, the spectrum of $H_0$ has a very rich structure in
terms of
the algebra parameters $\alpha_0$, $\alpha_1$, \ldots,~$\alpha_{\lambda-2}$.
This can be exploited to provide a bosonization of PSSQM of order $p =
\lambda-1$,
and, for $\lambda=3$, a bosonization of pseudoSSQM and OSSQM of order two.\par
%
%

\end{document}